# Analysis of Optimized Threshold with SLM based Blanking Non-Linearity for Impulsive Noise Reduction in Power Line Communication Systems

Ferheen Ayaz
School of Engineering and Informatics
University of Sussex
Brighton, BN1 9RH, UK
f.ayaz@sussex.ac.uk

Khaled Rabie, Bamidele Adebisi
School of Engineering
Manchester Metropolitan University
Manchester, M15 6BH, UK
k.rabie@mmu.ac.uk

*Abstract*—High amplitude impulsive noise (IN) occurrence over power line channels severely degrades the performance of Orthogonal Frequency Division Multiplexing (OFDM) systems. One of the simplest methods to reduce IN is to precede the OFDM demodulator with a blanking non-linearity processor. In this respect, Selective Mapping (SLM) applied to an OFDM signal before the transmitter does not only reduce Peak-to-Average Power Ratio (PAPR) but also increases the resulting Signal-to-Noise Ratio (SNR) when blanking non-linearity is applied at the receiver. This paper highlights another advantage of SLM based IN reduction, which is the reduced dependency on threshold used for blanking non-linearity. The simulation results show that the optimal threshold to achieve maximum SNR is found to be constant for phase vectors greater than or equal to 64 in the SLM scheme. If the optimized threshold calculation method is used, the output SNR with SLM OFDM will result in SNR gains of up to 8.6dB compared to the unmodified system, i.e. without implementing SLM. Moreover, by using SLM, we not only get the advantage of low peak power, but also the need to calculate optimized threshold is eliminated, thereby reducing the additional computation.

*Keywords—Blanking, impulsive noise, OFDM; peak-to-average power ratio (PAPR); power line communications (PLC); selective mapping (SLM); smart grid*

I. INTRODUCTION

Smart grid has been rapidly emerging as one of the biggest technological revolutions to the existing power grids and traditional energy resources [1-3]. It can be implemented by different technologies including power line communications (PLC). PLC is a well-suited networking technology to be used with smart grid because of its simple and easy deployment [4]. Moreover, PLC offers robust and reliable connectivity in environments such as underground, underwater and in building with walls made up of metals [5].

Multicarrier modulation schemes are used with PLC because of their characteristic of frequency selection [6]. A thoroughly investigated multicarrier modulation technique for PLC is Orthogonal Frequency Division Multiplexing (OFDM) [7]. OFDM has proved to be an effective solution to various problems associated with PLC like multipath and frequency selective attenuation [8-10]. However, despite of the robustness of OFDM, PLC still offers a number of challenges, including white noise and impulsive noise [11]. Among two categories of noise present in PLC channels, the impulsive noise (IN) plays a major role in overall performance degradation of the communication system [12]. In literature, asynchronous IN is defined as random short-lived pulse with high spectral density [13]. Commonly, it is analysed using Class-A Middleton noise model [14].

While it is impossible to remove IN completely from a signal, a number of methods for its elimination are present in the literature. Some methods remove IN at the cost of increased computation including multiple iterations [15-17] and some are based on intelligent but complex approaches such as fuzzy logic [18] and sparse Bayesian learning [8,19-20]. Simple and easy to implement non-linear pre-processors to remove IN are mainly based upon blanking [21] and clipping [22-24]. Blanking non-linearity is one of the most widely used methods which revolves around the concept of nulling the corrupted signals [25]. In this method, a block to nullify or blank the IN affected signals is introduced before an OFDM demodulator present at the receiver end. Some improved and modified versions of blanking are also present in literature to remove IN. Hybrid clipping-blanking [26], hybrid median-nulling [27], spectrum sensing, blanking and symbol retransmission [28] and SLM based blanking [29] are some of the examples.

In all blanking-based and most of other IN removal techniques, threshold selection is a critical parameter to maximize the performance. A threshold is defined as an amplitude level above which the OFDM signal is considered as corrupted by IN. Fixing a threshold value is crucial due to high Peak-to-Average Power Ratio (PAPR) of the OFDM signal. A very low value of threshold usually blanks the useful OFDM signals unaffected by impulsive noise and a high value of threshold is unable to detect IN-corrupted signals accurately [25]. Various threshold optimization techniques and adaptive threshold schemes are used to achieve optimum performance of IN removal techniques [30-32]. Some of these techniques are iterative [33] and require longer computational time. Moreover, techniques to reduce PAPR are also present in literature including clipping PAPR reduction, PAPR windowing and selective mapping (SLM) [34-35]. It has also been found that PAPR reduction performance is improved with the increasing number of phase vectors, thereby resulting in additional computation due to the increase in number of inverse fast Fourier transforms (IFFTs) required to produce the substitute signals [36].

In [30], an optimized threshold (OT) based on IN estimation through received signal characteristics; peak, median and mean is presented. This optimized threshold results in increased output Signal-to-Noise Ratio (SNR) for different IN probabilities, as compared to fixed threshold. It has already been shown that an optimized threshold can become independent of IN probability and initial estimation through PAPR reduction [29] and phase modulator transform system, referred as Constant Envelope OFDM (CE-OFDM) [37]. In this paper, we use the SLM technique, which already improves the blanking performance [29], and analyse its effect on the optimized threshold. The results of the study are in agreement with the conclusions presented in [29] and [37]. With the increase of phase vectors, the optimized threshold value remains constant regardless of the received signal characteristics.

The contribution of this work is twofold. Firstly, we will present a relation between optimized blanking threshold and the number of phase vectors. Secondly, the gain in SNR with SLM applied OFDM relative to the unmodified OFDM system is calculated. The results show that by the use of SLM, the additional computations to find adaptive or optimized threshold at the receiver end can be completely eliminated and a considerable gain in SNR is also achieved.

The organization of the rest of the paper is as follows. Section II describes the system model. Simulation results are analysed and discussed in Section III. Conclusions are presented in Section IV.

## II. SYSTEM MODEL

The system model used in this paper is illustrated in Fig. 1. The proposed model is a combination of models used in [29] and [30]. SLM block to reduce PAPR at the transmitter and Optimized Threshold Calculation Block at the receiver is introduced to the conventional blanking non-linearity based OFDM system. The transmitted OFDM signal $s(t)$ in time domain is obtained by taking Inverse Fourier Transform of the frequency domain signal and is given as

$$s(t) = \frac{1}{\sqrt{N}}\sum_{k=0}^{N-1} S_k \exp\left(j\frac{2\pi.k.t}{T_S}\right), 0 < t < T_s \quad (1)$$

where $S_k$ denotes the frequency-domain signal, $j=\sqrt{-1}$, $N$ is the number of subcarriers, and $T_s$ is the active symbol interval.

SLM technique generates set of $U$ different data blocks as

$$S = [S_0, S_1 \ldots\ldots\ldots S_{N-1}]^T \quad (2)$$

Each data block is multiplied by $U$ different phase vectors $W$ as

$$W = [W_o^u, W_1^u \ldots\ldots\ldots W_{N-1}^u]^T \quad (3)$$

where $u=1, 2, 3,\ldots\ldots\ldots\ldots U$. Inverse Discrete Fourier Transform of each block is obtained as

$$s^{(u)}(t) = \frac{1}{\sqrt{N}}\sum_{k=0}^{N-1-(u)} S_k \exp\left(j\frac{2\pi.k.t}{T_S}\right), 0 < t < T_s \quad (4)$$

where $s^{(u)}(t)$ and $\overline{S}_k^{(u)}$ are the SLM-OFDM signals in time and frequency domain, respectively.

The PAPR of the transmitted signal is given as

$$PAPR = \frac{\max|s(t)|^2}{E[|s(t)|^2]}. \quad (5)$$

The block with minimum PAPR is then selected for actual transmission and is found as

$$\bar{s}(t) = \arg\min_{0 \leq u \leq U-1}\{PAPR(s^{(u)}(t))\} \quad (6)$$

The transmitted signal is then passed through the PLC channel where Additive White Gaussian Noise (AWGN) and IN is added. AWGN is denoted by $w_k$ and has variance $\sigma_w^2 = (1/2)E[|w_k|^2]$ whereas the impulsive noise $i_k$ is modelled as Bernoulli-Gaussian random process [14] as follows

$$i_k = b_k g_k, \quad 0 \leq k \leq N-1 \quad (7)$$

where $b_k$ is the Bernoulli process of sequence '$b_k=1$' or '$b_k=0$' with probability of $p$ or $1-p$, respectively and $g_k$ is the complex–zero mean white Gaussian noise with variance $\sigma_i^2 = (1/2)E[|g_k|^2]$. Thus, the noisy channel can be characterized by the signal-to-background noise ratio $SBNR = 10\log_{10}(1/\sigma_w^2)$ and signal-to-IN ratio $SINR = 10\log_{10}(1/\sigma_i^2)$. The received time-domain signal $r_k$ can be expressed as

$$r_k = \begin{cases} s_k/\bar{s}_k + w_k & \text{if } b_k = 0 \\ s_k/\bar{s}_k + w_k + i_k & \text{if } b_k = 1 \end{cases}, k=0,1\ldots,N\text{-}1 \quad (8)$$

The output of the nonlinear pre-processor $y_k$ is then fed to the OFDM demodulator for the further processing and is represented in [21] as

$$y_k = \begin{cases} r_k & |r_k| \leq T \\ 0 & |r_k| > T \end{cases}, k = 0,1,\ldots\ldots N\text{-}1 \quad (9)$$

where $T$ is the blanking threshold. The performance of blanking non-linearity depends upon threshold value. An optimized threshold selection method given in [30] is used in our work which calculates the Optimized Threshold, $OT$ as

$$OT = \frac{INE}{\beta} \quad (10)$$

where $INE$ is the IN estimation given as

$$INE = \max(r_k) - \text{mean}(r_k) \quad (11)$$

and $\beta$ is

$$\beta = \gamma - \{\text{median}(r_k) - \text{mean}(r_k)\} \quad (12)$$

where $\gamma$ is a constant value. Note that we have set $\gamma = 7$ in throughout our simulations.

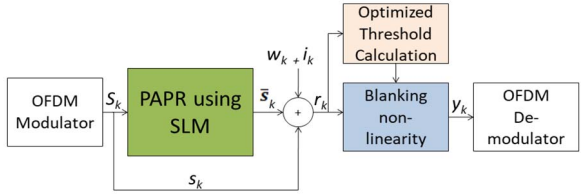

Fig. 1. Block Diagram of OFDM System with SLM at the transmitter and blanking with optimized threshold calculation at the receiver.

### III. SIMULATION RESULTS

The computer simulations using Matlab are carried out on OFDM system with N=64 subcarriers, 16 QAM modulation, SBNR = 40 dB, SINR = -10dB and $p$ = 0.01. The output SNR can be calculated as

$$SNR = \left(\frac{E[|s_k|^2]}{E[|y_k - s_k|^2]}\right) \quad (13)$$

### A. Optimized Blanking Threshold at U different phase vectors

Fig. 2 shows the optimized blanking threshold (OBT) at the maximum SNR obtained for blanking scheme with SLM at $U$ different phase vectors employed at the transmitter. A slight increase in OBT from $U$=1 to $U$=2 is observed. Then it gradually decreases from I=4 till $U$=64. At $U\geq$64, it becomes fixed regardless of the randomness of IN. This shows that increasing phase vectors makes the threshold independent of the noise and received signal characteristics thereby eliminating the need to compute $OT$ before passing the signal through blanking non-linearity. Maximum SNR can be achieved by blanking even with the fixed threshold.

### B. Analysis of OT used with SLM applied to OFDM signal

In [29] the output SNR at different blanking thresholds is calculated with various values of $U$. In our simulations, we first calculated $OT$ as per method defined in [30] and found output SNR only at $OT$ with various values of $U$. The simulations show that SLM applied to OFDM signal at the transmitter outperforms the blanking non-linearity used with unmodified OFDM signal, even if $OT$ is used as the threshold as in both cases. Figs. 3-5 explain the results.

In Fig. 3, the relative SNR gain/Loss is evaluated as

$$SNR_{\frac{Gain}{Loss}}(dB) = 10 \log_{10}(SNR_U/SNR_{unmod}) \quad (14)$$

Fig. 3 shows that at $U$=8, the loss in SNR of maximum -0.5dB only is observed at a point with respect to unmodified OFDM signal. However, at $U$>8, the SNR gain is positive at all points. The maximum gain is obtained at $U$=64.

The two major parameters; $\beta$ and $\Upsilon$ defined in [30] for calculation of $OT$ are also analyzed. Figs. 4 and 5 show that both result in better output SNR with SLM at the transmitter. However, the value of $\beta$ and $\Upsilon$ at maximum output SNR are increased with the increasing phase vectors. While $\Upsilon$ is a fixed parameter in [30] and $\beta$ is directly proportional to $\Upsilon$ as in (12), it implies that the probability of error in calculating $OT$ is still present as the maximum achievable output SNR depends upon the appropriate selection of $\Upsilon$.

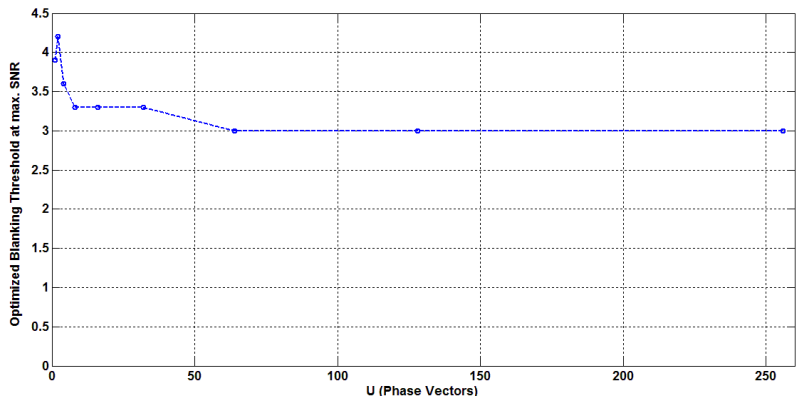

Fig. 2. Optimized Blanking Threshold found at maximum SNR for $U$=1 to 256.

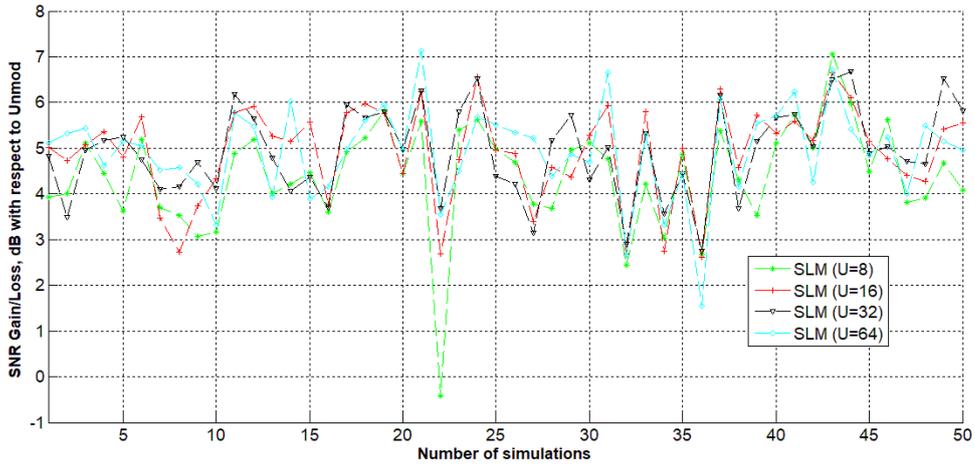

Fig. 3. Relative Gain/Loss by SLM applied to OFDM signal with unmodified OFDM signal using *OT* for 50 simulations.

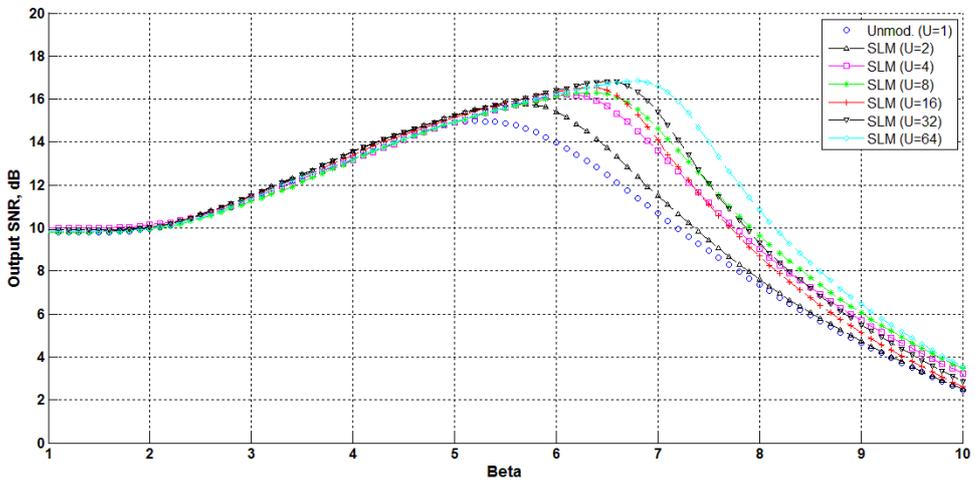

Fig. 4. SNR of blanking non-linearity with *OT* observed with respect to $\beta$.

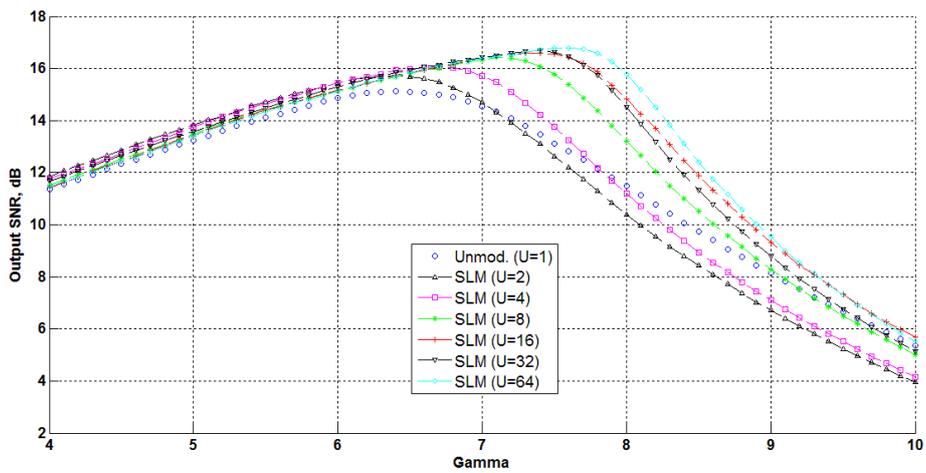

Fig. 5. SNR of blanking non-linearity with *OT* observed with respect to $\Upsilon$.

## IV. CONCLUSION

The optimized threshold calculation method is analysed in this paper. It is found that the need for threshold calculation at receiver end is eliminated if SLM is applied at transmitter with large number of phase vectors ($U \geq 64$) and a fixed optimum threshold can be used. This may increase the computational complexity at the transmitter end but can reduce calculation at the receiver side. However, at the nominal number of phase vector (($U \geq 8$), the performance of optimized threshold calculation method is enhanced and an increase of up to 7dB can be obtained in the output SNR. Since the optimized threshold calculation method is simply based upon statistical characteristics of received signal only, a high output SNR is achievable by using SLM with less number of phase vectors at the transmitter and an optimized threshold calculation block at the receiver. This will result in performance enhancement with reduced peak power and less complexity. For further investigations, another PAPR reduction technique can be used in place of SLM to observe its effect on blanking non-linearity and threshold optimization.

## REFERENCES

[1] V.C. Gungor, D. Sahin, T. Kocak, S. Ergut, C. Buccella, C. Cecati, G. P. Hancke, "A survey on smart grid potential applications and communication requirements," IEEE Transactions on Industrial Informatics (2013) pp. 28-42.

[2] Z. Fan, P. Kulkarni, S. Gormus, C. Efthymiou, G. Kalogridis, M. Sooriyabandara, Z. Zhu, S. Lambotharan,, W. H. Chin, "Smart grid communications: Overview of research challenges, solutions, and standardization activities," IEEE Communications Surveys & Tutorials 15(1) (2013) pp. 21-38.

[3] B. Adebisi, K. M. Rabie, A. Ikpehai, C. Soltanpur, A. Wells, "Vector OFDM transmission over non-Gaussian power line communication channels," IEEE Systems Journal, (2017)pp. 1-9.

[4] S. G. Yoon, S. Jang, Y. H. Kim, S. Bahk, "Opportunistic routing for smart grid with power line communication access networks," IEEE Transactions on Smart Grid 5(1) (2014) pp. 303-311.

[5] X. Cheng, R. Cao, L. Yang, "Relay-aided amplify-and-forward powerline communications," IEEE Transactions on Smart Grid 4(1) (2013) pp. 265-272.

[6] A. M. Tonello, F. Versolatto, M. Girotto, Multitechnology "(I-UWB and OFDM) coexistent communications on the power delivery network," IEEE Transactions on Power Delivery 28(4) (2013) pp. 2039-2047.

[7] M. V. Ribeiro, G. R. Colen, F. V. De Campos, Z. Quan, H. V. Poor, "Clustered-orthogonal frequency division multiplexing for power line communication: when is it beneficial?," IET Communications 8(13) (2014) pp. 2336-2347.

[8] J. Lin, M. Nassar, B. L. Evans," Impulsive noise mitigation in powerline communications using sparse Bayesian learning," IEEE Journal on Selected Areas in Communications 31(7) (2013) pp. 1172-1183.

[9] E. S. Hassan "Multi user MIMO-OFDM-based power line communication structure with hardware impairments and crosstalk," IET Communications 11(9) (2017) pp. 1466-1476.

[10] X. Liu, J. Lou, H. Sun, H. Liu, X. Gu, "A timing synchronization method for OFDM based power line communication" 9th International Conference in Communication Software and Networks (ICCSN) IEEE (2017) pp. 543-547.

[11] A. Mathur, M. R. Bhatnagar, B. K. Panigrahi, "Performance evaluation of PLC under the combined effect of background and impulsive noises," IEEE Communications Letters 19(7) (2015) pp. 1117-1120.

[12] M. Zimmermann, K. Dostert, "Analysis and modeling of impulsive noise in broad-band powerline communications," IEEE transactions on Electromagnetic compatibility 44(1) (2002) pp. 249-258.

[13] R. Barazideh, B. Natarajan, A. V. Nikitin, R. L. Davidchack, "Performance of analog nonlinear filtering for impulsive noise mitigation in OFDM-based PLC systems, "IEEE 9th Latin-American Conference on Communications (LATINCOM) (2017) pp. 1-6.

[14] D. Middleton, "Canonical and quasi-canonical probability models of class A interference," IEEE Transactions on Electromagnetic Compatibility (2) (1983) pp. 76-106.

[15] J. Häring, A. H. Vinck, "OFDM transmission corrupted by impulsive noise," Proc. Int. Symp. Powerline Communications (ISPLC), (2000) pp. 9-14.

[16] J. Häring, A. H. Vinck, "Iterative decoding of codes over complex numbers for impulsive noise channels," IEEE Transactions on Information Theory 49(5) (2003) pp. 1251-1260.

[17] R. Liu, T. L. Kung, K. K. Parhi, "Impulse noise correction in OFDM systems," Journal of Signal Processing Systems 74(2) (2014) pp. 245-262.

[18] M. V. Ribeiro, C. A. Duque, J. M. T. Romano, "An interconnected type-1 fuzzy algorithm for impulsive noise cancellation in multicarrier-based power line communication systems," IEEE Journal on Selected Areas in Communications 24(7) (2006) pp. 1364-1376.

[19] J. Lin, B. L. Evans, "Non-parametric mitigation of periodic impulsive noise in narrowband powerline communications," Global Communications Conference (GLOBECOM) IEEE (2013) pp. 2981-2986.

[20] J. Lin, M. Nassar, B. L. Evans, "Non-parametric impulsive noise mitigation in OFDM systems using sparse Bayesian learning," Global Communications Conference (GLOBECOM) IEEE (2011) pp. 1-5.

[21] S. V. Zhidkov, "On the analysis of OFDM receiver with blanking nonlinearity in impulsive noise channels," Intelligent Signal Processing and Communication Systems IEEE (2004) pp. 492-496.

[22] H. A. Suraweera, C.Chai, J. Shentu, J. Armstrong, "Analysis of impulsive noise mitigation techniques for digital television systems," Research Paper, Department of Electronic Engineering, La Trobe Univeristy, 2003.

[23] F. H. Juwono, Q. Guo, D. Huang, K. P. Wong, "Deep clipping for impulsive noise mitigation in OFDM-based power-line communications," IEEE Transactions on Power Delivery 29(3) (2014) pp. 1335-1343.

[24] F. H. Juwono, Q. Guo, D. Huang, K. P. Wong, "Joint peak amplitude and impulsive noise clippings in OFDM-based power line communications," 19th Asia-Pacific Conference on Communications (2013) pp. 567-571.

[25] S. V. Zhidkov, "Performance analysis and optimization of OFDM receiver with blanking nonlinearity in impulsive noise environment," IEEE Transactions on Vehicular Technology 55(1) (2006) pp. 234-242.

[26] S. V. Zhidkov, "Analysis and comparison of several simple impulsive noise mitigation schemes for OFDM receivers," IEEE Transactions on Communications 56(1), 2008.

[27] Z. Ali, "Hybrid median-nulling scheme for impulsive noise mitigation in OFDM transmission.," Fourth International Conference on In Aerospace Science and Engineering (ICASE) IEEE (2015) pp. 1-5.

[28] G. Bartoli, R. Fantacci, D. Marabissi, L. Micciullo, D. Tarchi, "Detection and mitigation of impulsive interference on OFDM signals based on spectrum sensing, blanking and

symbol retransmission," Wireless personal communications 77(4) (2014) pp. 2631-2647.
[29] K. M. Rabie, E. Alsusa, "Efficient SLM based impulsive noise reduction in powerline OFDM communication systems," Global Communications Conference (GLOBECOM) IEEE (2013) pp. 2915-2920.
[30] Z. Ali, F. Ayaz, C. S. Park, "Optimized threshold calculation for blanking nonlinearity at OFDM receivers based on impulsive noise estimation," EURASIP Journal on Wireless Communications and Networking (2015) pp. 1-8.
[31] U. Epple, M. Schnell, "Adaptive threshold optimization for a blanking nonlinearity in OFDM receivers," Global Communications Conference (GLOBECOM) IEEE (2012) pp. 3661-3666.
[32] E. Alsusa, K. M. Rabie, "Dynamic peak-based threshold estimation method for mitigating impulsive noise in power-line communication systems," IEEE Transactions on Power Delivery 28(4) (2013) pp. 2201-2208.
[33] M. R. Ahadiat, P. Azmi, A. Haghbin, "Impulsive noise estimation and suppression in OFDM systems over in home power line channels," International Journal of Communication Systems, 30(1) (2017)
[34] J. Singh, R. Garg, I. K. Aulakh,, "Comparison of PAPR reduction techniques in OFDM based Cognitive Radio," International Journal of Applied Engineering Research 11(5) (2016).
[35] M. V. Malode, D. B. Patil, "PAPR reduction using modified selective mapping technique," Int. J. of Advanced Networking and Applications 2(02) (2010) pp. 626-630.
[36] S. J. Heo, H. S. Joo, J. S. No, D. W. Lim, & D. J. Shin, "Analysis of PAPR reduction performance of SLM schemes with correlated phase vectors," IEEE International Symposium on Information Theory, (June, 2009) pp. 1540-1543.
[37] K. M. Rabie, E. Alsusa, A. D. Familua, L. Cheng," Constant envelope OFDM transmission over impulsive noise power-line communication channels," International Symposium on Power Line Communications and its Applications (ISPLC) IEEE (2015) pp. 13-18.